\documentclass[preprint2]{aastex}

\shorttitle{SPECTROPOLARIMETRY OF R CrB IN 1998--2003}
\shortauthors{Kawabata et al.}

\begin{document}

\title{SPECTROPOLARIMETRY OF R CORONAE BOREALIS IN 1998--2003: 
DISCOVERY OF TRANSIENT POLARIZATION AT MAXIMUM BRIGHTNESS}

\author{K.~S.~Kawabata\altaffilmark{1,2},
Y.~Ikeda\altaffilmark{3,2},
H.~Akitaya\altaffilmark{4,2},
M.~Isogai\altaffilmark{1,2},
K.~Matsuda\altaffilmark{5},
M.~Matsumura\altaffilmark{6},
O.~Nagae\altaffilmark{7}, and
M.~Seki\altaffilmark{8}
}

\altaffiltext{1}{Hiroshima Astrophysical Science Center,
 Hiroshima University, 1-3-1 Kagamiyama, Higashi-Hiroshima,
 Hiroshima 739-8526, Japan; 
 kawabtkj@hiroshima-u.ac.jp, isogai@hiroshima-u.ac.jp}

\altaffiltext{2}{Visiting Astronomer, Okayama Astrophysical
 Observatory of National Astronomical Observatory of Japan (NAOJ)}

\altaffiltext{3}{Photocoding, Higashi-Hashimoto 3-16-8-101, Sagamihara,
 Kanagawa 229-1104, Japan;
 ikeda@photocoding.com}

\altaffiltext{4}{Division of Optical and Infrared Astronomy, 
 NAOJ, Osawa 2-21-1, Mitaka, Tokyo 181-8588, Japan;
 akitaya@optik.nao.ac.jp}

\altaffiltext{5}{Nishi-Harima Astronomical Observatory, Sayo-cho,
 Hyogo 679-5313, Japan; 
 matsuda@nhao.go.jp}

\altaffiltext{6}{Faculty of Education, Kagawa University,
 Saiwai-cho 1-1, Takamatsu 760-8522, Japan; 
 matsu@ed.kagawa-u.ac.jp}

\altaffiltext{7}{Department of Physical Science, School of Science, 
 Hiroshima University, 1-3-1 Kagamiyama, Higashi-Hiroshima, 
 Hiroshima 739-8526, Japan;
 nagae@hep01.hepl.hiroshima-u.ac.jp}

\altaffiltext{8}{Astronomical Institute, Graduate School of Science,
 Tohoku University, Aramaki, Aoba-ku, Sendai 980-8578, Japan;
 seki@astr.tohoku.ac.jp}

\begin{abstract}
We present an extended optical spectropolarimetry of R CrB
from 1998 January to 2003 September.
The polarization was almost constant in the phase of maximum 
brightness, being consistent with past observations.
We detected, however, temporal changes of polarization ($\sim 0.5$ \%)
in 2001 March and August, which were the first detection of 
large polarization variability in R CrB near maximum brightness.
The amplitude and the position angle of the `transient polarization' 
were almost constant with wavelength in both two events.
There was a difference by about 20 degrees in the position angle 
between the two events.
Each event could be explained by light scattering due to 
short-lived dust puff occasionally ejected off the line of sight.
The flatness of the polarization against the wavelength 
suggests that the scatterer is a mixture of dust grains 
having various sizes.
The rapid growth and fading of the transient polarization 
favors the phenomenological model of dust formation near the 
stellar photosphere (e.g., within two stellar radii)
proposed for the time evolution of brightness 
and chromospheric emission lines during deeply 
declining periods, although the fading timescale can hardly
be explained by a simple dispersal of expanding dust puff
with a velocity of $\sim 200-350$ km s $^{-1}$.
Higher expansion velocity or some mechanism to destroy the
dust grains should be needed.
\end{abstract}


\section{INTRODUCTION}

R Coronae Borealis (RCB) stars are hydrogen-deficient,
carbon-rich variables that undergo declines in visual
brightness up to 7 mag or more at irregular intervals.
The evolutionary pathways of these stars are still unclear,
with the white dwarf merger model and the final 
helium-shell flash model being suggested for the
extreme abundances (e.g., \citealt{ibe96,sai02,cla07}).
The darkening phenomenon has been attributed to eclipse by clouds
of dust grains (dust puff) formed along the line of sight
\citep{lor34,oke39}.
Near-infrared excess, attributable to the dust emission
at an equivalent blackbody temperature of $\sim 900$ K,
was found for the prototype of this group, R CrB \citep{ste69},
and this excess is a general property of RCB stars \citep{fea73}.
The infrared radiation shows variations on timescales of 
$\sim 3.5$ yr \citep{fea97b} and does not clearly correlate 
with the optical brightness \citep{for72,fea97b,yud02}.
These indicate that a large amount of dust being produced
surrounds the star and that a smaller dust puff is formed
in a random direction per any one ejection event 
(e.g., \citealt{ohn01,ohn03}).
Recent near-infrared adaptive optics observations with an 
8-m class telescope directly revealed the presence of multiple 
dusty clouds in the vicinity of RY Sgr \citep{lav04}.

The dust formation mechanism itself is, however, still one of the
fundamental problems in RCB stars (e.g., \citealt{cla96,fea97a,fea00}).
\citet{fad86} derived from a theoretical argument
a conclusion that the dust formation has to occur somewhere
$\sim 20 R_{\rm *}$ apart from the central star
where the local temperature becomes low enough to allow for
carbon grains to condense.
On the other hand, observational facts such as the time 
evolution of chromospheric emission lines and timescales 
of brightness recovering from declines are favorable to
other models in which dust forms in close proximity 
($\lesssim 2 R_{\rm *}$) to the central star \citep{cla92,whi93}.
To settle this discrepancy, some mechanism to promote
dust formation near RCB stars have been proposed.
\citet{woi96} suggested that the local gas temperature can be
substantially decreased by supercooling after the passage
of a shock wave driven by stellar pulsation.
\citet{asp96} proposed an episodic local density 
inversions in the ionization layer of helium
due to a kind of radiative instability which
can promote a blob ejection.
\citet{fea97a} discussed the dust formation above the cool 
regions of giant convection cells.
\citet{sok99} suggest that cool magnetic spots existing 
inside of photosphere facilitate dust formation
above the spots after a shock wave passes.

Polarimetry is a probe sensitive to circumstellar medium
near the stellar surface, and even to dust formation episodes 
around RCB stars on very small spatial scales.
If dust forms actually near the photosphere 
($\lesssim 2 R_{\rm *}$), the angular size of the dust puff 
should be more than $50\arcdeg$ for total eclipse of the stellar disk.
Such a wide dust puff is likely to scatter the light from the star
and may cause a net polarization more than $0.5$--$1$ \%,
when it locates along the direction nearly orthogonal to
the line of sight \citep{cod95}.
Therefore, it is natural to expect that RCB stars temporally show a
significant polarization at maximum brightness stage.
The wavelength dependence of the polarization would give  
unique information about the ``newborn'' dust grain.
So far only small fluctuation of intrinsic polarization 
less than 0.2 \% has been found for RCBs near maximum brightness
\citep{ser69,coy73,coy74,efi80,efi90,sta88,ros90,whi92a,tra94,
cla95,cla97,yud03,bie06},
except for \citet{efi80}'s observation on 1974 August 24 
($p_{\rm V} = 0.38\pm 0.03$ \%).

\begin{figure}
\plotone{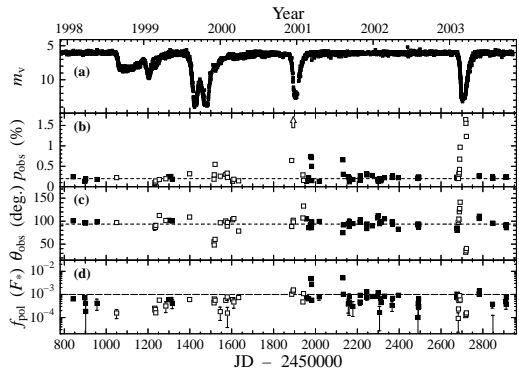}
\caption{Time variation of polarimetric properties of R CrB.
This figure shows (a) visual magnitude from VSNET, 
(b) observed polarization amplitude, 
(c) its position angle, and (d) polarized flux at synthetic $V$ band 
against Julian date.
The visual magnitude data are from VSNET database.
Polarization data are shown by filled squares for observations
at maximum brightness stage ($\Delta m_{\rm V} \leq 0.3$) 
and by open squares during declines 
($\Delta m_{\rm V} > 0.3$) , with observational error ($1\sigma$).
The horizontal lines in (b) and (c) denote the constant component of 
polarization at $V$ band (see \S 3.2).
The arrow in (b) indicates that there are data points
much larger than the shown range.
The polarized flux is derived from variable component of
polarization and magnitude drop (see \S 3.3).
In the maximum brightness stage R CrB generally showed a 
nearly constant polarization, while it showed significant 
temporal variation in 2001 March and August.
}
\label{fig1}
\end{figure}

In this paper, we present a new spectropolarimetry of R CrB
on 84 nights in 1998--2003.
We successfully found two events of temporal increase in
polarization by about 0.5 \%\ at maximum brightness stage.
We consider the period when the visual magnitude is within 
0.3 mag below the maximum $V$ magnitude ($V_{\rm max}$) as
``maximum brightness'' phase and call the other period 
``decline'' phase.
We adopt $V_{\rm max} = 6.09$ which is the mean magnitude
during the apparent visual maximum of R CrB,
JD2,450,800--51,000 and 52,000--52,650 (Figure \ref{fig1}a).
The photometric data referred in this paper are obtained from 
the VSNET\footnote{Variable Star NETwork; 
\url{\tt http://www.kusastro.kyoto-u.ac.jp/vsnet/}} 
and AAVSO\footnote{American Association of Variable Star
Observers; \url{\tt http://www.aavso.org/}} databases.

\section{OBSERVATIONS AND DATA REDUCTION}

The data were obtained with the low-resolution spectropolarimeter,
HBS (an acronym of ``Henkou-Bunkou-Sokkou-Ki'' which
stands for spectropolarimeter in Japanese; \citealt{kaw99}) 
from 1998 January 28 to 2003 September 29 at the Dodaira Observatory 
and the Okayama Astrophysical Observatory of NAOJ.
At the Dodaira Observatory, HBS was attached to the Cassegrain focus of
the 0.91 m telescope ($F/18$; $12\farcs 4$ mm$^{-1}$ at the focal plane).
At Okayama Astrophysical Observatory, HBS was attached to the
Cassegrain focus of either the 0.91 m telescope 
($F/13$; $17\farcs 2$ mm$^{-1}$) or the 1.88 m
telescope ($F/18$; $6\farcs 1$ mm$^{-1}$).
HBS has a superachromatic half-wave plate and a quarts Wollaston
prism; the orthogonally polarized spectra are simultaneously
recorded on either a front-illuminated type TI CCD 
(1024$\times$1024 pixels, 12 $\micron$ square per pixel) 
or a back-illuminated type SITe CCD (512$\times$512 pixels, 
24 $\micron$ square per pixel).
We used either the 1.4 mm$\phi$ circular hole (D1) or the
0.2 mm $\times$ 1.4 mm rectangular hole (D2) as a focal
diaphragm.
Each diaphragm has two holes of the same dimension, and we put a
target star in one hole and the nearby sky in the other.
For the observations with D1 diaphragm, the spectral resolution
depends on the seeing size because the stellar image size is
much smaller than the diameter of the circular hole.
For the observations with D2, the spectral resolution also
depends on the stellar image size at the Okayama 0.91 m telescope,
while it does not at the other telescopes because the stellar 
image size was usually larger than the narrower dimension of 
the rectangular hole.
Nevertheless, the spectral resolution falls in the range between 
50\ \AA\ and 150\ \AA\ in the all observations.

A unit of the observing sequence consisted of successive
integrations at four position angles (PA), $0\arcdeg$, $22\fdg 5$,
$45\arcdeg$, and $67\fdg 5$, of the half-wave plate.
The obtained images were processed using the reduction package
for HBS data, which was outlined in \citet{kaw99}.
Instrumental polarization was derived from unpolarized 
standard star data obtained almost in each night.
The data were averaged in each observation run (typically 
$\sim 10$ days for the Okayama 1.88 m telescope and 1-2 months for
the other telescopes). The level of instrumental 
polarization $p_{\rm instr}$ was well expressed by a 
smooth function of wavelength, and vectorially removed
from observed Stokes $Q$ and $U$ spectra.
At any run, the stability ($1\sigma$) of $p_{\rm instr}$
was less than 0.05 \%.
The factor of instrumental depolarization was obtained from 
observation through a Glan-Taylor prism.
This observation also gave us the wavelength-dependent 
PA of the equivalent optical axis of the
superachromatic half-wave plate.
The zero point of the PA on the sky was determined
from observations of strongly polarized standard stars
listed in \citet{wol96}.
The instrumental polarization, the instrumental depolarization, 
and the zero point of the PA were properly corrected for 
in the reduction package.

The observations are summarized in Table \ref{tbl-1}, in which
polarization data integrated over the synthetic 
Johnson {\it V} band filter \citep{bes90} are shown.
For the observations with D2 diaphragm, we also show the 
equivalent widths of the C$_{2} (0-0)$ 5165 \AA\ band in the table.
The flux was calibrated using spectrophotometry of 
some standard stars \citep{tay84}.

\section{RESULTS}

\subsection{Overview}

Figure \ref{fig1} shows the time variation of polarization at
the synthetic {\it V} band, together with the visual light curve.
R CrB experienced three major photometric declines 
in our observation period.
It demonstrates large polarization variability during declines;
for example, the polarization angle rotated by more than
30$\arcdeg$ in 1999 December and the polarization amplitude 
became over 5 \% in 2000 December.
On the other hand, R CrB near maximum brightness showed
nearly constant polarization (except for on 2001 March 9--12 
and August 9).
These characteristics are consistent with past observations 
(e.g., \citealt{efi80,sta88,cla95}).
A commonly-accepted explanation for the large polarization during 
decline is that the unpolarized direct stellar light is heavily
obscured by the dust cloud (dust puff) along the line of sight 
and the visible flux is predominantly scattered light.

\begin{figure}
\plotone{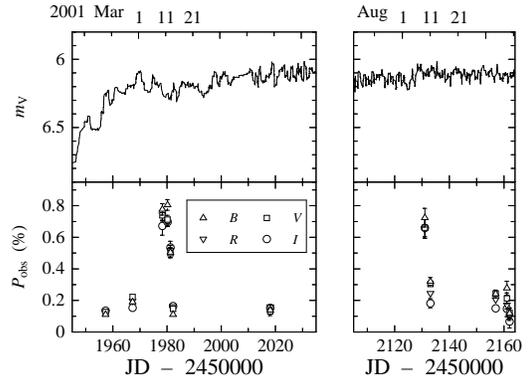}
\caption{Time variation of visual magnitude and
observed polarization in the epochs including 
the transient polarization events.
For the lightcurve we show smoothed (17 point running-mean)
AAVSO validated magnitudes, which includes a wealth of data
points (about four times more than VSNET ones).
Triangles, squares, inverse triangles, and circles denote
the amplitude of observed polarization in synthetic $B$, $V$, $R$,
and $I$ bands, respectively.
It seems that there is temporal photometric variations
of $\sim$0.1 mag a few days prior to the polarization peaks,
which could be connected with small dust formation episodes.
}
\label{fig2}
\end{figure}

\begin{figure}
\plotone{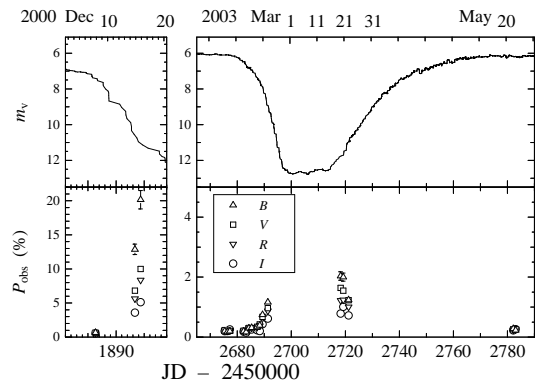}
\caption{Same as Figure 2 during 2000-2001 and 2003 declines.
The data are plotted with the same manner in Figure 2.
The amplitude of polarization increases as the star fades.
Bluer polarization increases more rapidly than red one, which
is consistent with the past polarimetry.
The ratio of polarization amplitude to $\Delta m_{V}$
was higher in 2000--2001 decline than in 2003 decline.
}
\label{fig3}
\end{figure}

\begin{figure}
\plotone{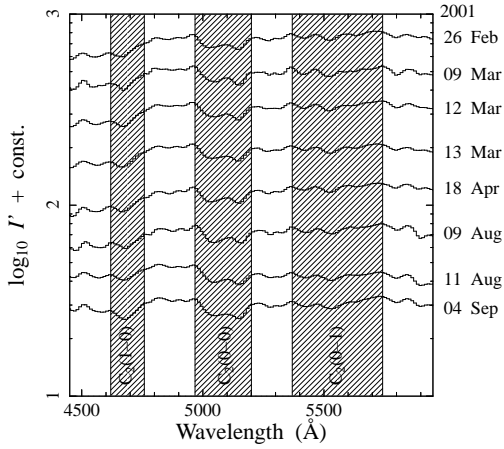}
\caption{Sample spectra of R CrB in the maximum brightness
stage including the periods of the transient polarization.
The system response (instrument$+$telescope) is corrected but 
airmass effect is not corrected in those spectra.
The date of observation is indicated at right side of each spectrum. 
Hatched area denotes the molecular bands (C$_{2}$).
No significant variation is seen in the period.
}
\label{fig4}
\end{figure}

However, we detected temporal increase of 
polarization two times even at maximum brightness stages; 
the first event was observed 
on 2001 March 9--12 and the second was on 2001 August 9.
In both times, the amplitude of polarization increased by 
$\Delta p\simeq 0.5$ \% (Figure \ref{fig2}).
Although polarization usually increases accompanied with 
the onset of major decline phases (Figure \ref{fig3}),
we cannot see any significant decline ($\Delta V\gtrsim 0.3$) 
in the lightcurve around the two events.
Also, any remarkable change in flux spectra have not been
recognized as shown in Figure \ref{fig4}, and the equivalent width 
of C$_{2}\ (0-0)$ 5165 \AA\ absorption band (Table \ref{tbl-1}) 
kept an almost constant value, 22--26 \AA , which is 
consistent with the typical value of 
R CrB near maximum brightness (17.5--43 \AA\ with an 
average of 26.5 \AA ; \citealt{cla95}).
It should be noted, however, that there is possible 
photometric variations accompanied with the polarimetric
variations.
The lightcurve shows possible shallow dips ($\sim$0.1 mag) a 
few days prior to the polarization events (Figure \ref{fig2}), 
which could be connected with small dust formation episodes.
We will discuss this temporal polarimetric activities
after subtraction of the constant polarization component.

\subsection{Constant Component of Polarization}

It has been known that R CrB shows a nearly constant 
polarization at maximum brightness.
\citet{sta88} averaged the polarization observed in a period 
when no brightness fluctuations were seen, and derived a
constant component of $Q=-0.21$ \%, $U=-0.07$ \%
($p=0.22$ \%, $\theta=99\arcdeg$) as the values
averaged over the optical wavelengths.
The same values were obtained from averaging subsequent
non-decline observations in 1990--1993 \citep{cla95}.
\citet{cla97} reported that the observation in 1995 March,
before the onset of the decline of 1995, showed a typical
Serkowski-type ISP function ($p_{\rm max}=0.24\pm 0.08$ \%,
$\lambda_{\rm max}=0.46\pm 0.01$ $\micron$, $\theta\sim 100\arcdeg$)
which was in good agreement with the previous estimations.
\citet{efi80} assumed the interstellar component of 
$p_{\rm max}=0.2$ \% and $\theta = 110\arcdeg$ from polarization 
of nearby field stars.
Similar technique also gives $p_{\rm max}=0.20$ \% and
$\theta = 97\arcdeg$ \citep{bie06}.

We averaged our polarization data obtained on 46 nights when R CrB was
at maximum brightness stage ($\Delta m_{\rm V} \leq 0.3$ mag)
except those in the periods of the transient polarization.
The mean values for Stokes $Q$ and $U$ parameters and their standard
deviations in each band are
$(Q_{\rm B}, U_{\rm B}) = ( -0.198\pm 0.068\ \% , -0.020\pm 0.055\ \% )$,
$(Q_{\rm V}, U_{\rm V}) = ( -0.203\pm 0.049\ \% , -0.035\pm 0.053\ \% )$,
$(Q_{\rm R}, U_{\rm R}) = ( -0.188\pm 0.044\ \% , -0.028\pm 0.056\ \% )$, and
$(Q_{\rm I}, U_{\rm I}) = ( -0.161\pm 0.044\ \% , -0.026\pm 0.055\ \% )$.
Parameters of Serkowski function \citep{ser75,whi92b} corresponding
to those values are derived by a non-linear regression as
$p_{\rm max}=0.204\pm 0.046$ \%,
$\lambda_{\rm max}=0.48\pm 0.27$ $\micron$ and
$\theta=94\fdg 1 \pm 4\fdg 1$.
These are consistent with the constant and/or interstellar 
component derived in previous studies.
The constancy of the component over two decades, being much more 
than the $\sim$3.5 yr IR period, allows us to
consider that the component is fully interstellar polarization.
It is noted that \citet{cla95} found a small variability of 
polarization up to 0.14 \%\ (possibly correlated with the light curve,
at least, for their 1991 data) during the maximum brightness stage.
The variability seems rather random, and such an effect would be 
cancelled out in our averaged values, while they possibly remain 
in the standard deviations.

\section{DISCUSSION}

After vectorial subtraction of the constant component from the
observations, we obtain variable components of polarization,
$\ p_{\rm var}(\lambda )$ and $\theta_{\rm var}(\lambda)$.
Their values in synthetic $V$ band are shown in Table \ref{tbl-1}.

It is known that polarized flux of R CrB is relatively constant
since 1968 at the level of $\sim 10^{-3}$ to $10^{-4}\ F_{\rm *}$
at any part of declines \citep{whi92a,cla95,cla97}, 
where $F_{\rm *}$ is the total stellar flux at maximum brightness stage.
Here, the polarized flux is simply derived as
\begin{equation}
 f_{\rm pol} = p_{\rm var} \times 10^{-0.4 \Delta m_{\rm V}} 
\ F_{\rm *}\mbox{ ,}
\end{equation}
where $\Delta m_{\rm V}$ is the drop of visual 
magnitude from maximum $V_{\rm max}=6.09$ (see \S~1).
In our data (Figure \ref{fig1}d), $f_{\rm pol}$ also falls between 
$10^{-3}$ and $10^{-4}\ F_{\rm *}$ on almost all days both during 
and out of declines.
However, at the epochs of the transient polarization,
$f_{\rm pol}$ showed $\sim$5 times as much as the typical
maximum values, which indicates that the net polarization
flux was extraordinarily large.

\subsection{Origin of the Transient Polarization}

\begin{figure}
\plotone{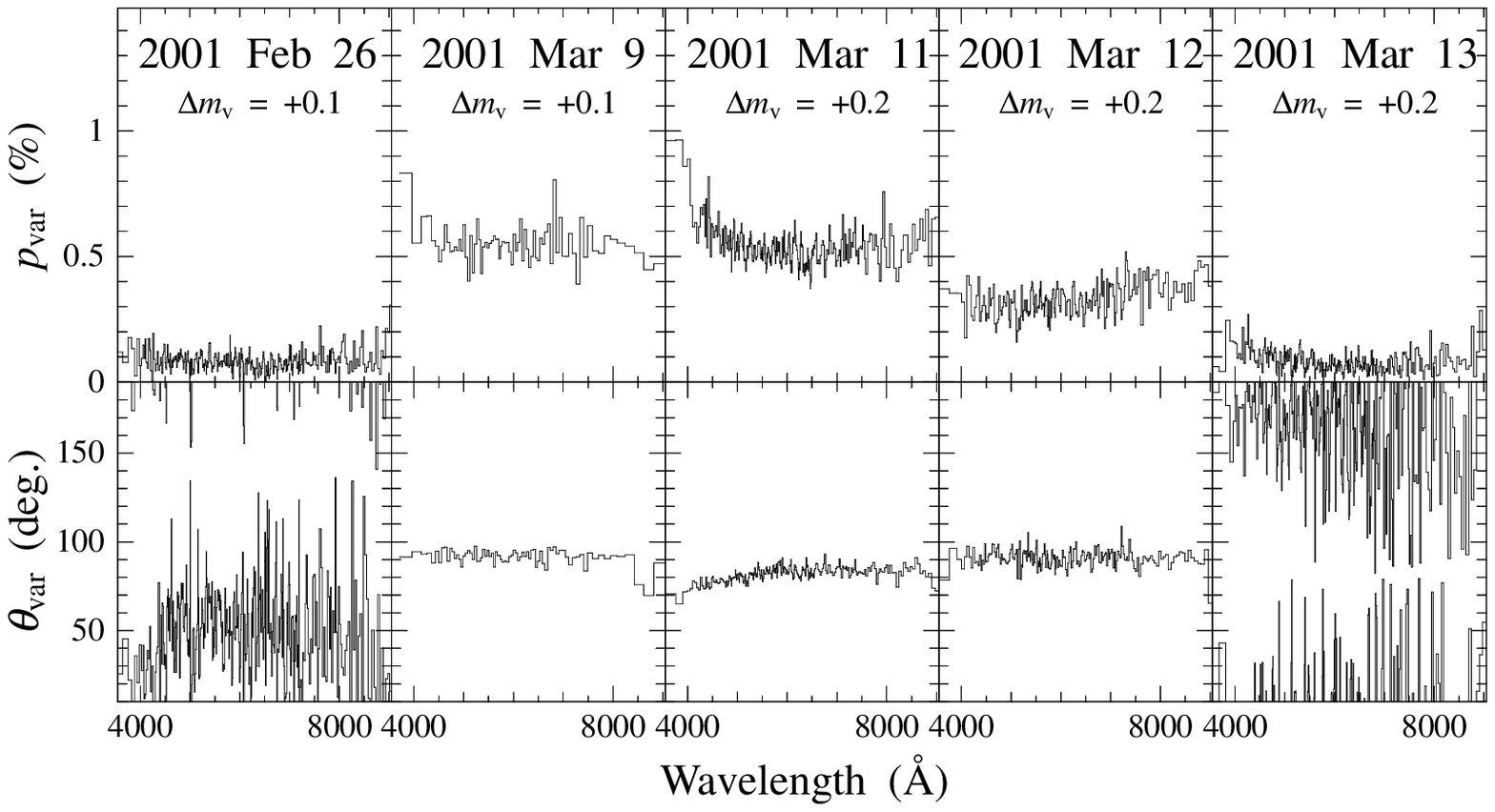}
\plotone{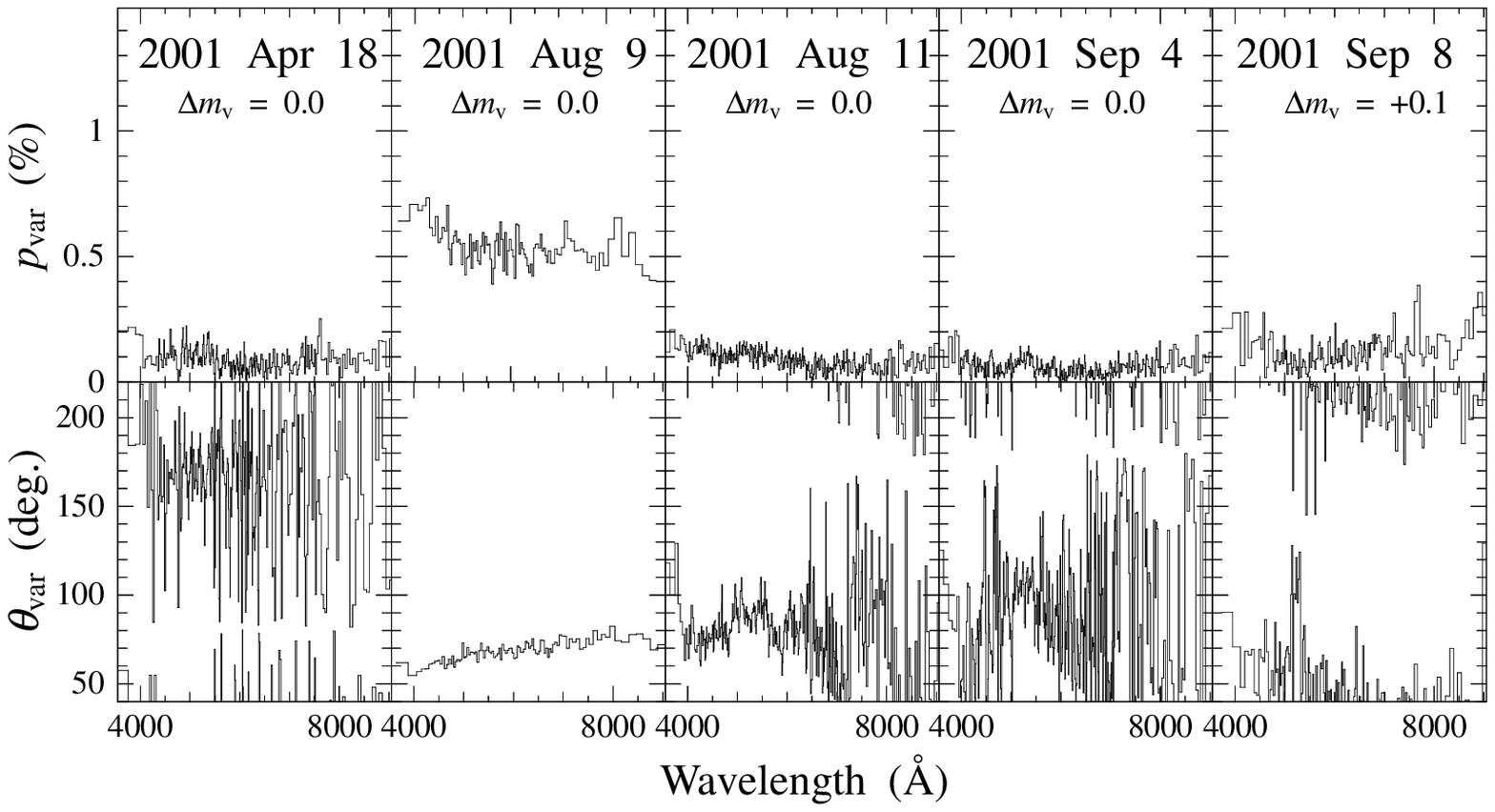}
\caption{Temporal variation in the wavelength dependence 
of the variable component of polarization at maximum 
brightness phases: Upper combined figure shows polarization
amplitude and its position angle around the first event 
of the transient polarization (2001 March) and the lower one 
shows the second one (2001 August). 
The data are binned to a constant photon noise of 0.04\%.
The date of observation and the $V$ magnitude below the 
maximum $m_{V}=6.09$ are indicated in each upper panel.
}
\label{fig5}
\end{figure}

\begin{figure}
\epsscale{.70}
\plotone{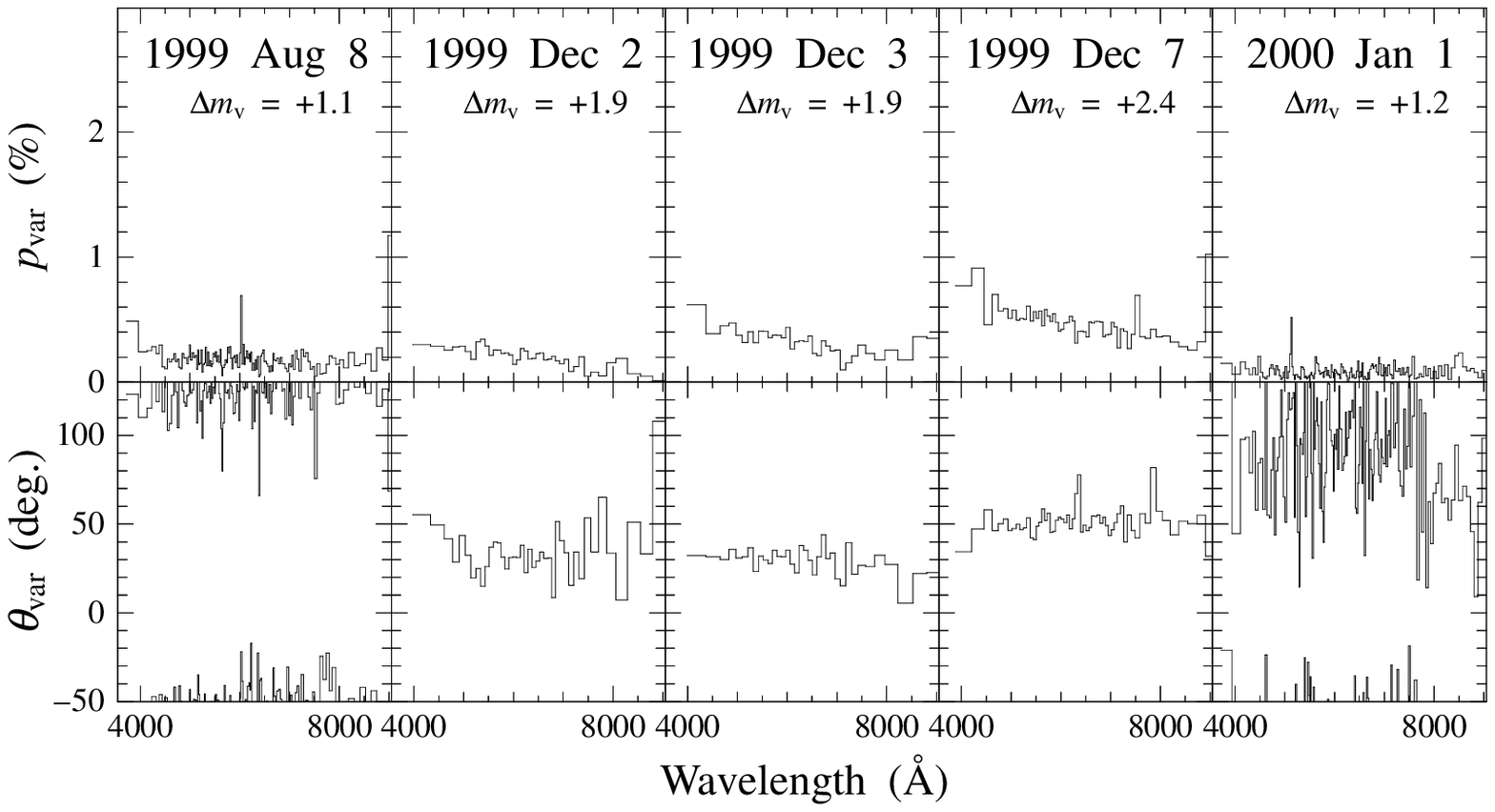}
\plotone{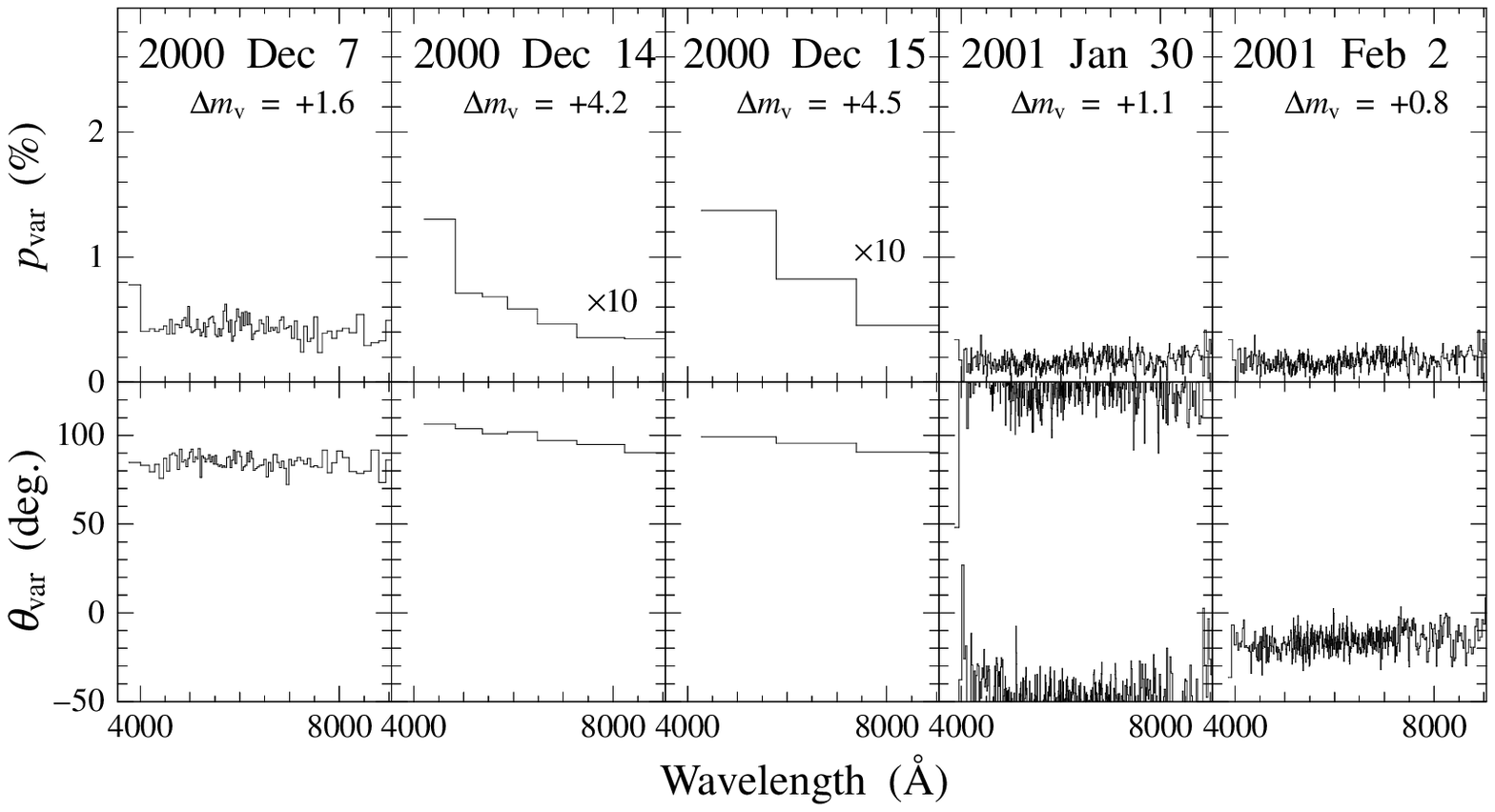}
\plotone{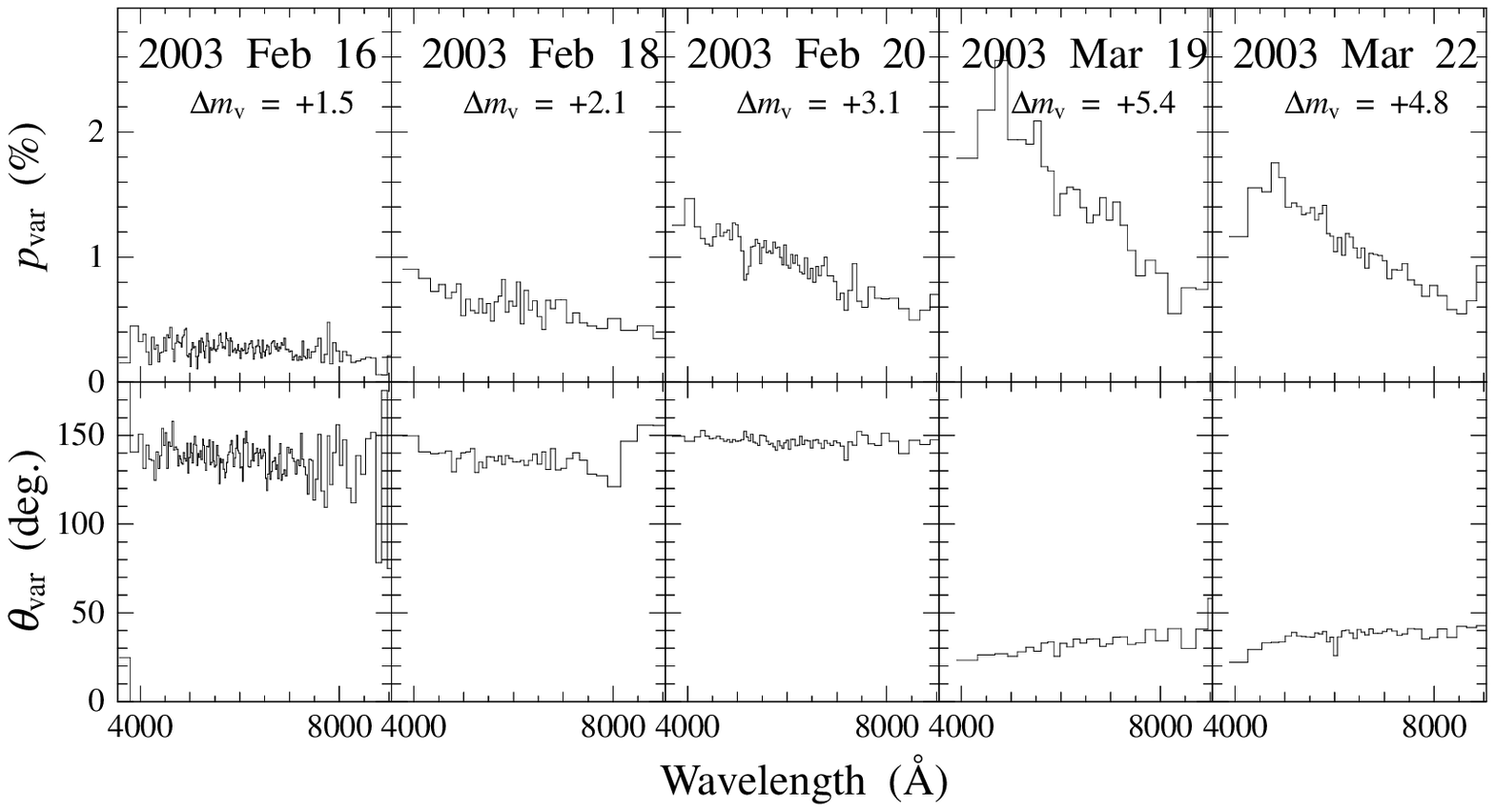}
\caption{Same as Figure 5 during decline phases.
The polarization during decline phase generally 
has a wavelength dependence; $p_{\rm var}$ is larger
at bluer wavelengths.
}
\label{fig6}
\end{figure}

Our polarimetric monitoring brought first detection of the
`transient polarization' amounting to 0.5 \% for R CrB in 
the maximum brightness phase.
As far as we know, there exists only one observation in which
R CrB near maximum brightness showed polarimetric variation larger 
than $0.2$ \%: \citet{efi80} found $\Delta p = 0.38\pm 0.03$ \%\ 
at $\lambda_{\rm eff}=$5450 \AA\ on 1974 August 24 $=$ 
JD$2,442,284$.
Figures \ref{fig5}a and \ref{fig5}b show the time variation of
$p_{\rm var}(\lambda)$ and $\theta_{\rm var}(\lambda)$ 
in the periods including the first and second events of the 
transient polarization.
There is no significant wavelength dependence 
in position angle for both events, compared with 
the polarization found during decline phase 
(Figures \ref{fig6}a--c).

Suppose that several clouds exist around R CrB and that they 
scatter the light from the star.
If each cloud was ejected from the photosphere in a random manner
and it contains a different polarizing property from one to another, 
the position angle of the scattered light would be mostly 
wavelength dependent.
The nearly constant $\theta$ against $\lambda$ indicates 
that the polarization can be `effectively' attributed 
to scattering by one component.
The wavelength dependence of $p_{\rm var}(\lambda)$ was 
almost flat in optical wavelengths
and quite similar between the two cases.
For the event on 1974 August 24, $p(\lambda )$
seems also flat (0.38--0.46 \%) in the wavelength range,
3630--7450 \AA\ \citep{efi80}.
The similarity in $p_{\rm var}(\lambda )$ curve suggests
an existence of the possible generality in the polarizing mechanism 
among those three events.
Rayleigh scattering by large molecules or smallest solid particles 
($\lesssim \lambda / 10$) cannot reproduce such wavelength dependence 
because the scattering coefficient is highly wavelength-dependent
($\propto \lambda^{-4}$).
Scattering by free electrons can reproduce a flat 
$p_{\rm var}(\lambda )$ curve.
However, it is unlikely for R CrB to have a large amount of 
ionized gas which can produce an observable polarization
because the effective temperature of R CrB is not so high 
($T_{\rm eff}\simeq 6900$ K; \citealt{sch75,asp97}).
The emission lines which appears in R CrB during declines
(e.g., \citealt{rao99,rao06}) are not so strong that spectra
at maximum brightness stage have little emission line.

As described in \S 1, several observational facts suggest 
that R CrB intermittently ejects dust puffs even at 
maximum brightness phases.
The transient polarization could be naturally explained by a 
dust puff ejection off the line of sight.
The shallow dips in the lightcurve (Figure \ref{fig2}) would 
support this idea, if we can assume that the dust puff eclipses
a small part of the photosphere at the very early (nearest) stage.
Single-sized grain of radius $a \simeq 0.1\ \micron$ shows a peak 
polarization at optical wavelength and the peak wavelength 
varies with the grain radius (e.g., \citealt{sha75}).
For flat $p_{\rm var}(\lambda)$ curve, the dust grains would have
a size distribution of a wide range of $a$.

\begin{figure}
\epsscale{.70}
\plotone{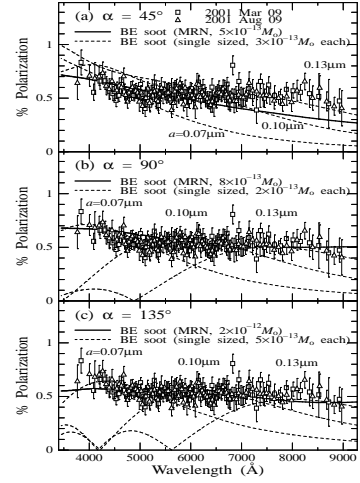}
\caption{Samples of wavelength dependence of polarization
calculated with Mie scattering theory.
From top to bottom, we show them in the case of the scattering 
angle $\alpha=$45$\arcdeg$, 90$\arcdeg$ and 135$\arcdeg$,
respectively.
The dust model is an amorphous carbon (BE soot; \citealt{rou91}).
Open squares and triangles are variable components of polarization
observed on 2001 March 9 and on August 9, respectively.
Dashed lines denote the results for scattering by single-sized 
($a=0.07,\ 0.10,\ 0.13$ \micron ) grains.
Thick solid lines are results for a mixture of grains of an 
inverse power-law size distribution ($n(a)\propto a^{-3.5}$) 
having lower and upper cutoffs at 0.005 $\micron$ and 
0.25 $\micron$, respectively \citep{mat77}.
In this model we assume an optically-thin case and the 
the dust mass corresponds to $r=2R_{\rm *}$ indicated in the 
parentheses should be considered as approximated values. 
}
\label{fig7}
\end{figure}

We performed a calculation of Mie scattering with a simplified 
model consisting of a point-like dust cloud and a central star,
and estimated the light from the stellar system; that is the sum
of an unpolarized direct stellar light and the 
partially-polarized scattered light.
In the calculation we assumed that the dust grains are all 
spherical amorphous carbon (e.g., \citealt{lam01}) and adopted
the refractive indices of the BE soot \citep{rou91}.
We also assumed that the cloud is optically-thin, i.e., multiple 
scattering process is negligible.
We set the distance of the cloud $r=2R_{\rm *}$ where $R_{\rm *}$ 
is the stellar radius ($= 85 R_{\sun}$; \citealt{fea75}) and
make an estimation of the total mass of dust necessary for
explaining the $0.5$ \% polarization.
The results are shown in Figure \ref{fig7}.
In the case of forward-scattering (scattering angle 
$\alpha < 90\arcdeg$, where $\alpha$ is the angle between  
the direction of the incident light and that of the 
scattered one),
the $p(\lambda )$ curves gradually decrease with wavelength
and they do not fit well the observation.
For a right angle scattering and backward-scattering 
($\alpha > 90\arcdeg$) cases,
the single-sized grain model fails to reproduce 
the flat $p(\lambda)$ curve. The size distribution 
of the carbon dust should range over $\sim$0.13 $\micron$
in order to produce the flat $p_{\rm var}(\lambda)$ curve.
We here consider a power-law size distribution 
like a well-known interstellar dust model 
(\citealt{mat77}; hereafter, MRN).
The MRN size distribution is expressed as $n(a)\propto a^{-\beta}$ 
within $a_{\rm min} \leq a \leq a_{\rm max}$, where 
$\beta = 3.5,\ a_{\rm min}=0.005\ \micron$ and 
$a_{\rm max}= 0.25\ \micron$.
Results of the MRN model seem not to contradict 
the observation in a wide range of scattering
angle ($\alpha=60\arcdeg$--$135\arcdeg$).
This is only an illustrative example.
A large choice of parameters 
($\alpha,\ a_{\rm min},\ a_{\rm max},\ \beta$) can yield 
similar $p_{\rm var}(\lambda )$ curves.
In general, the scattering by grains having 
various sizes seems a plausible mechanism for the transient 
polarization.
The similarity in the observed $p(\lambda )$ curves between/among 
two (or three, including \citealt{efi80}'s event) events suggest 
that the dependence of $p(\lambda )$ on the scattering angle
is not so strong.

A correlation between pulsational phase and the time of
decline onset has been found for RY Sgr and V854 Cen which show 
fairly regular pulsation cycles than R CrB \citep{pug77,law92}.
Although the correlation is less significant for R CrB itself
\citep{pug77,gon83,per87,law91,fer94}, it is recently confirmed
by \citet{cra07}.
These suggest that the condensation of dust is triggered 
by the stellar pulsation.
According to this picture, the appearance of the transient
polarization is likely to correlate with the pulsation phase.
In our observation, the interval between the two polarization
events was $151\pm 1$ d.
It has been known that R CrB shows a semi-regular pulsation
of about 40 d from photometric and spectroscopic 
observations, for example $38.6$ d \citep{ale72}, 
$P=39.8$ d \citep{yud02} and $P=42.7$ d \citep{rao99}.
If we assume that the 151 d interval corresponds to four 
cycles, the length of one cycle is $\sim 38$ d.
This seems a little shorter compared with the pulsation period.
However, it is noted again that the pulsation period of R CrB is 
not coherent. Some shorter periods have been found in a timescale of 
a few hundred days.
\citet{fer94} found well-developed oscillations with an
apparent period of $36.7\pm 0.9$ d in late 1992 in the visual
light curve; while they also found that the better-defined
peaks observed from 1992 through 1993 could be fitted with 
a single, linear ephemeris of period $35.3\pm 0.2$ d.
We cannot exclude the possibility that the transient
polarization was correlated with the stellar pulsation.

\subsection{Size Distribution of Dust}

In the previous section, we showed that the wavelength dependence
of the transient polarization can be reproduced by light scattering
due to the carbon grain with various sizes including grains larger 
than $\sim$0.13 \micron.
However, UV extinction properties at the 1980 decline of R CrB
indicate that the dust was grassy or amorphous carbon grain 
with a size distribution of $a=0.005$--$0.06\ \micron$ \citep{hec84}.
A detailed analysis of extinction curves \citep{zub97}
for three observations of R CrB during declines in 1980--1984
indicated that the size distribution of graphite or amorphous 
carbon grains has a distinct peak-like form (widths of 
$\Delta a=0.002$--$0.006\ \micron$) with typical sizes
of $a = 0.03$ through $0.07\ \micron$.
If such small particles scatter the light from the central star,
the $p_{\rm var}(\lambda)$ curve should show a slope of
decreasing with wavelength in the optical (Figure \ref{fig7}).

The amplitude of polarization is proportional to inverse square 
of the distance between the central star and the dust puff, $r$,
while the extinction depends only on the column density
along the line of sight.
The information derived from the extinction would be
a mean property of the dust clouds along the line of sight
toward the star, while the information
derived from the polarization would be selectively for the dust 
close to the central star.
Therefore the apparent difference in dust size between
the estimate from polarimetry and that from UV extinction
may indicate that the size distribution of the dust changes 
with the distance from the central star.
\citet{coy73} reported the variability of the wavelength
dependence of polarization in 1972 decline of R CrB, and 
interpreted each observation with scattering by graphite dust 
of different sizes ranging from $0.05$ to $0.10\micron$ in radius.
In their results the mean particle size became smaller 
unsteadily with time.
\citet{eva85} proposed that the extinction
properties of the dust responsible for 1982 decline of the
hot RCB star, MV Sgr, can be explained by somewhat large 
carbon grains ($a\sim 0.2\ \micron$).
Another model applied for the same extinction data also 
gave a similar result \citep{zub97}.
For some extinction data of other RCB stars, he also suggested 
that the particle size exceeds $a=0.1\ \micron$.
These facts suggest that the scattering matter consist of
discrete clouds with different mean particle sizes 
(depending on their ages and their distance from the star)
even in the same star.
If the dust forms near the photosphere,
a grain evaporation mechanism might play a role on
the change of the size distribution (e.g., \citealt{eva93}).
The particle size inferred from our polarimetry
suggests that the evolutionary phase of the dust grains
causing the transient polarization is more or less the same
from that of the dust grains obscuring the stellar flux
during deep declines.


\subsection{On the Short Timescale of Fading}

The transient polarization in 2001 March 
diminished gradually between March 11 and 13.
For the second event, the polarization also diminished within two days.
It is remarkable that the decaying timescale of the transient 
polarization is only $\sim 2$ days, which is much shorter 
than the recovery times of the brightness from deep minimum 
(typically several weeks).
If we assume that the scattering geometry keeps similar, 
the polarized flux approximately depends on the product of 
intensity of the stellar light at the scattering material
and total scattering cross section. 
Thus, the amplitude of polarization decreases as the dust cloud
goes away from the star (i.e., larger $r$) or the dust particles 
are destroyed (i.e., smaller $a$).

Here we consider the former case. If the decrease of polarization 
was only due to radial dispersal of dust cloud, the rate of 
decrease in polarization roughly depends on the radial distance $r$ 
and the mean expansion velocity of the cloud $v_{\rm exp}$.
The distance can be expressed as $r=R_{\rm df}+v_{\rm exp}t$, 
where $R_{\rm df}$ is the distance between the central star
and the place where the dust formation finishes, and $t$ is 
the time elapsed from the cessation of the dust formation.
For simplicity, we assume that the scattering cloud is optically thin.
In this case, the polarization amplitude approximately changes as 
$p(r) = p_{\rm c} D(r)(R_{\rm *}/r)^{2}$,
where $p_{\rm c}$ is a constant and 
$D(r)=\{1-(R_{\rm *}/r)^{2}\}^{0.5}$ is the 
finite disk depolarization factor \citep{cas87}.\footnote{The 
optical thickness of the cloud might not be 
practically negligible, which results in depolarization.
As the cloud expands, the cloud becomes 
optically-thin and the polarizing efficiency gradually recovers.
Thus, $p$ should be damped more slowly than $D(r)\cdot r^{-2}$ in the 
vicinity of the star and we obtain the lower-limit of 
$v_{\rm exp}$ with this method anyhow.}
This function monotonically decreases with $r$ for 
$r>\sqrt{3/2}R_{\rm *}$, and the most rapidly decreasing 
case occurs for nearly $R_{\rm df}=\sqrt{3/2}R_{\rm *}$.
The observed transient polarization decreased to 
one fourth within $\sim 2$ days.
These indicate that the dust cloud needs to travel, 
at least, $1.9 R_{\rm *}$ within two days, i.e.,
$v_{\rm exp} \gtrsim 650$ km s$^{-1}$.
Thus, if the expansion velocity exceeds $650$ km s$^{-1}$,
the dispersal of dust cloud can explain the timescale 
of the polarization extinction.
However, the derived velocity is not consistent with the 
typical mass loss velocity observed in R CrB.
The absorption component of the \ion{Na}{1} D or the
\ion{He}{1} $\lambda$10830 line are blueshifted by up to 
$-150$ to $-350$ km s$^{-1}$ 
(e.g., \citealt{pay63,que78,cot90,rao99,cla03}).
Evolutions of emission line system at the onset of 
declines suggest that the wind velocity is likely to be lower 
($\sim 50$ km s$^{-1}$) at close proximity to 
the photosphere, $\simeq 1$--$2 R_{\rm *}$
\citep{pay63,ale72,cla96}.
Although higher-velocity wind have been reported in 
the unusual RCB stars, V854 Cen (400--1000 km s$^{-1}$, 
\citealt{cla93,rao93}), the observational facts
suggests the ejecta of R CrB is gradually 
accelerated up to $\sim 200$ km s$^{-1}$ within a
few weeks (probably by radiation pressure on dust grains).
It is unlikely that the dust puff is accelerated 
to $v_{\rm exp} \gtrsim 650$ km s$^{-1}$ within 
$R\sim 1.2R_{\rm *}$.


Therefore, the rate of depolarization can hardly be explained 
only by the radial expansion of the dust puff.
If the dust forms actually near the photosphere 
($\lesssim 2 R_{\rm *}$), the strong radiation from the 
central star can evaporate and destroy the dust grains once the
possible cooling instability to promote the dust condensation
(see \S 1) terminates.
In any case, the rapid disappearance of polarization 
prefers that the dust grains are formed in an activated region, 
that is, a region close to the photosphere.

The problem of dust formation in such hostile environments is
also under discussion in the field of Wolf-Rayet stars.
Some phenomenological similarities 
with RCB stars have been reported for some late-type WC 
stars and [WC] PN nuclei (e.g., \citealt{wil97,cro97,vee98,cla01,kat02}).
Polarization is relatively sensitive to the region close
to the light source for some optically-thin systems,
and the transient polarization events represent 
the potential of polarimetry as the probe to the
dust formation around such stars.

\subsection{Geometry of Mass Loss}

\begin{figure}
\plotone{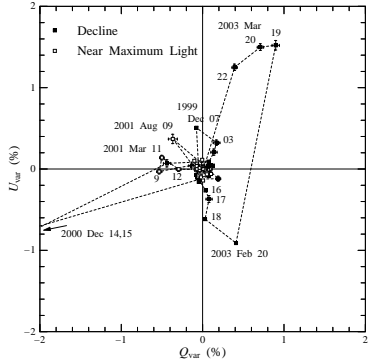}
\caption{Variation of {\it V} band polarization in $QU$ diagram. 
The constant component of polarization has been subtracted. 
The symbols show observations at nearly maximum brightness
(filled squares) and those during declines (open square).
Error bar represents $1\sigma$ observational error of 
individual observation. The mass ejection events in 1999-2003
do not seem to follow the bipolar model suggested by 
\citet{cla97} from past observations during decline phases, 
in which the data points lie at PA=0--30\arcdeg\ and 90--120\arcdeg\ on 
the projected sky (corresponding to PA=0--60\arcdeg\ and 
180--240\arcdeg\ in this diagram).
}
\label{fig8}
\end{figure}

\citet{sta88} presented that the PA of polarization in R CrB 
observed during decline of 1986 was along with data of two
previous declines of 1968 and 1972 \citep{ser69,coy73}.
This suggests that there was a preferred plane 
(PA $\simeq 30$--$45\arcdeg$) in the dust ejection.
However, the PAs observed during 1989 and 1993 declines 
were significantly different from them \citep{cla95}.
\citet{cla97} summarized a data set in five declines 
with polarimetry (from 1971 through 1996) having wavelength
information when R CrB was more than 5 mag below maximum 
brightness, and presented that they well fit to the 
two opposite sector regions of PA $=15\arcdeg\pm 15\arcdeg$ and 
$=105\arcdeg\pm 15\arcdeg$ in $QU$ diagram.
They suggested that dust ejection in R CrB has
a bipolar geometry consisting of obscuring torus
and bipolar lobes.
We can see a similar tendency in our data at decline 
phases (Figure \ref{fig8}).
The PA of the largest polarization found during 
the 2000 decline ($90\arcdeg$--$100\arcdeg$) falls
in the sector regions.
The PA in the recovering phase of the 2003 decline 
($30\arcdeg$--$36\arcdeg$) also falls near the sector regions.
These implies that the bipolar geometry for dust ejection
directions still exists in R CrB.
It is noted that the transient polarization was at
$80\arcdeg$--$90\arcdeg$ (2001 March) and $\sim 70\arcdeg$ 
(2001 August), which lies near or only slightly apart from 
the edge of the sector regions.

During deep declines a considerable rotation ($\gtrsim 40\arcdeg$) 
of the PA with wavelength has been occasionally found in
some RCB stars (e.g., \citealt{whi92a,cla97}).
This can be explained by an optical depth effect where a
thick torus obscures the central star and thin bipolar
lobes act as a reflection nebula \citep{cla97}.
On the other hand, for the transient polarization, 
the rotation of the PA with wavelength is small.
It supports the idea that only a single dust puff, which is 
closest to the photosphere, contributes the transient 
polarization effectively. The size distribution discussed 
in \S 4.2 would reflect the real properties of the newborn 
grains selectively.

\citet{ohn01} carried out {\it K}-band speckle interferometric
observations of R CrB on 1996 October 1 ($V=7$) and 1999 September
28 ($V=10.61$) and suggested that the large dust shell
(with an inner boundary at $r=82 R_{\rm *}$ and 920 K) has a
rather isotropic distribution, at least, in a scale of larger
than $\sim 300 R_{\rm *}$.
The authors also introduced a newly formed dust cloud close to
the star ($r=4.5 R_{\rm *}$ and 1200 K) to fit both the visibility 
and the spectral energy distribution simultaneously.
\citet{asp97} gave an analysis of SED with a model atmosphere and 
suggested that the IR excess could be explained by recently formed, 
hot dust grains ($\sim 2000$K) amounting to a total mass of 
$\sim 10^{-11}\ M_{\sun}$.
Some extinction models for RCB stars suggested that
the total mass (gas$+$dust) per ejection event were
$10^{-9}$--$10^{-7}\ M_{\sun}$ \citep{fea86,fad88,cla92}
although these could be affected by uncertainties in $r$,
$\phi$ (opening angle of the dust puff), and the gas-to-dust ratio.
\citet{yud02} derived a mean dust formation rate of
$3.1\times 10^{-9}\ M_{\sun}$ yr$^{-1}$ with a radiative
transfer modeling.
Our Monte Carlo simulation (see Appendix) suggests that 
the dust puff off the line of sight can produce a polarization
of $\simeq 0.5$\% when it has a mass of 
$\gtrsim 9\times 10^{-13}\ M_{\sun}$ and an opening angle of 
$\gtrsim 40\arcdeg$ for $r\sim 2 R_{\rm *}$.
The actual mass should be highly model dependent.
However, the lower-limit of the dust mass is still much less than 
the mean dust formation rate ($\sim 10^{-9}\ M_{\sun}$ yr$^{-1}$),
and we conclude that the amplitude of polarization in the 
transient polarization events would be consistent with 
the random ejection model in which dust puffs are formed 
in the close vicinity of the stellar atmosphere.

\section{SUMMARY AND CONCLUSION}

We carried out spectropolarimetric monitoring of R CrB
from 1998 January to 2003 September.
The constant component of polarization derived by \citet{sta88} 
and confirmed by \citet{cla95} still existed in our observational period.
In addition, we discovered a temporal increase of polarization up to 
$\simeq 0.5$ \% at maximum brightness stage in 2001 March and August.
It is likely that dust puff which was ejected off the line of sight 
produced the transient polarization.
Shallow declines ($\sim$0.1 mag) were marginally found a few days
prior to the polarization events, which would support this idea
if we can assume that the dust puff eclipsed a part of the photosphere
at the very early stage.
The flatness in $p_{\rm var}(\lambda)$ curve of the transient 
polarization, which was a common property for both two cases, 
suggested that the size distribution of the dust was
not like a $\delta$-function (as the results of UV--optical 
extinction studies) but rather wide, e.g., a well-known 
model of interstellar dust (MRN size distribution).
We suggest that the size distribution evolves 
with the age or the distance from the star.
The fading timescale ($\sim 2\pm 0.5$ days) of the polarization 
is fairly rapid compared with the recovering rate of brightness from 
deep declines.
It cannot be interpreted by a simple dissipation 
due to radial expansion of the dust puff at the 
expansion velocity of $\sim 200-350$ km s$^{-1}$, 
even in the case where the dust formation 
occurred at $r \sim 2\ R_{\rm *}$.
Higher expansion velocity or some mechanism to
destroy the dust grains should be needed.

So far we have not detected a possible increasing
part of the transient polarization, which would be
important to discuss the process and the site of 
the dust formation around RCB stars.
Further spectropolarimetric monitoring of RCB stars near 
maximum brightness will bring a valuable information of a 
dust formation and destruction processes in a hostile 
circumstellar environment.

\acknowledgments

We thank G.~C.~Clayton, the referee, for helpful suggestions.
We are grateful to T.~Cho,
H.~Iwamatsu, S.~Hamasaka, N.~Hirakata, R.~Hirata,
K.~Homma, K.~Hoshino, T.~Karube, N.~Kobayashi,
M.~Kondoh, S.~Masuda, Y.~Nakamura, S.~Nakayama,
A.~Okazaki, I.~Ota, M.~Saito, T.~Sakurai, R.~Suzuki, 
C.~Tokoku, K.~Yoshioka 
for their kind supports to the observation.
We also thank to the staff members of Dodaira Observatory
and Okayama Astrophysical Observatory of NAOJ for technical
supports, especially to Y. Norimoto, H. Shibasaki and K. Okita.
We are indebted to the variable star observations from the
VSNET and AAVSO database contributed by observers worldwide
for the light curve of R CrB.
Data reduction/analysis was in part carried out on the ``sb'' computer
system operated by the Astronomical Data Analysis Center (ADAC)
and Subaru telescope of NAOJ.
NAOJ is an interuniversity research institute of astronomy
operated by the Ministry of Education, Culture, Sports, Science and 
Technology.
This work was supported by a Grant-in-Aid from the Ministry
of Education, Culture, Sports, Science and Technology of
Japan (No. 17684004).

\appendix

\section{MONTE-CARLO SIMULATION OF POLARIZATION BY DUST PUFF}

Monte Carlo method is a useful tool to solve
problems including multiple scattering of light
since it requires only a few assumptions and it can be applied 
for many kinds of complicated geometries and particles.
The details in application of Monte Carlo method for 
light scattering process is described in, e.g.,
\citet{war83}, and \citet{hil91}.

\begin{figure}
\plotone{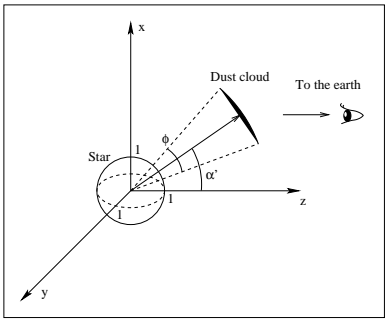}
\caption{Geometry of the scattering model in Monte Carlo simulation.
Light source ($=$ the central star) is located at the origin.
The scattering cloud is a part of a spherical shell
centered at the origin with an opening angle of $\phi$
and having a radius of $r$.
The center of the cloud is on the $x$-$z$ plane. 
The direction of the earth is that of $z$-axis.
}
\label{fig9}
\end{figure}

As seen in Figure \ref{fig9},
we consider the Cartesian coordinate, (x, y, z).
We define that the origin is the center of the star.
The stellar radius is 1, i.e., the coordinate is normalized 
by the stellar radius.
The direction of the earth is {\boldmath $e$}$=(0,0,1)$.
The shape of the dust puff is a part of geometrically-thin spherical shell 
with a radius of $r$ and an opening angle of $\phi$ from the origin.
The center of the dust puff, {\boldmath $r$}, is located 
in the $x$-$z$ plane.
These enable us to investigate the depolarization effect 
due to finite solid angle of the photosphere in the neighborhood of the
star, which is a different point from \citet{cod95}'s.
Optical depth of the cloud along {\boldmath $r$} is $\tau _{0}$.
For simplicity, we neglect geometrical thickness of the shell.
The scattering angle, that is, the angle between 
{\boldmath $r$} and {\boldmath $e$}, is $\alpha$.

In our Monte Carlo simulation,
more than a million photons are emitted from random sites
on the stellar surface toward random directions.
All released photons have the same intensity,
which means that rim darkening effect is neglected.
When they meet the cloud, some penetrate,
some are absorbed by, and the others are scattered by 
dust grains in the cloud.
The scattered photons are partially polarized and emitted again
toward a new direction according to a phase function 
{\boldmath $R$}$(\alpha )$.
We adopt a single peaked Henyey-Greenstein function \citep{whi79}
as the phase function:
\begin{equation}
\mbox{\boldmath $R$} (\alpha) = \frac{3}{4} \left(
\begin{array}{cccc}
P_1 & P_2 & 0   & 0 \\
P_2 & P_1 & 0   & 0 \\
0   & 0   & P_3 & -P_4 \\
0   & 0   & P_4 & P_3
\end{array}
\right),
\end{equation}
and,
\begin{eqnarray}
P_1 &=& \frac{1-g^2}{(1+g^2-2g\cos \alpha)^{3/2}}, \\
P_2 &=& -p_{\rm l} \frac{1-\cos ^2 \alpha}{1+\cos ^2 \alpha} P_{1}, \\
P_3 &=& \frac{2\cos \alpha}{1+\cos ^2 \alpha} P_{1}, \\
P_4 &=& -p_{\rm c}\frac{1-\cos ^2 \alpha _{\rm f}}{1+\cos ^2 
\alpha _{\rm f}} P_{1}, \\
\alpha _{\rm f} &=& \alpha (1+3.13s \exp [-7\alpha / \pi]),
\end{eqnarray}
where $g$ is an asymmetric parameter
(ranging from 0 for isotropic scattering to 1 for forward-throwing),
$p_{\rm l}$ and $p_{\rm c}$ are the maximum linear and circular 
polarization, respectively,
and $s$ is the skew factor ($=1$ in almost all cases). 
Three parameters $\alpha$, $g$,
and $p_{\rm l}$, are related with the linear polarization.
$g$ varies with a wavelength and a mean size of dust.
\citet{whi79} investigated the dependency of $g$ on wavelength 
for a well-known model of interstellar graphite dust.
We used his values, lacking a better alternative for 
a mixture of cosmic grains having various sizes.
Referring to Figure 2 of his paper, we adopt $g=0.46$ at 5500 \AA. 
We also obtain the albedo $\omega = 0.61$ at 5500 \AA\ from 
the same figure.
The $p_{\rm l}$ is set to 0.51 \citep{cod95}.
Although the single peaked Henyey-Greenstein phase function is 
not well approximated at longer wavelength region and for 
back-scattering cases \citep{whi79}, 
the difference can be negligible in our cases, 
i.e., $x = 2\pi a / \lambda 
\lesssim 1$ and $\alpha \leq 135\arcdeg$.
More details of our calculation will be described in a
preparing paper \citep{ike07}.

\subsection{Size and Mass of Dust Puff}

\begin{figure}
\plotone{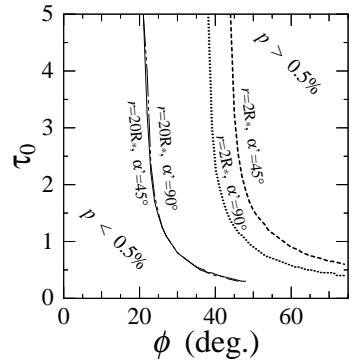}
\caption{Relations between the opening angle 
$\phi$ and optical thickness $\tau_{0}$ satisfying $p=0.5$ \%.
Thick dashed line is for $r=2 R_{*}$ and $\alpha'= 45 ^{\circ}$,
thick dotted one is for $r=2 R_{*}$ and $\alpha'= 90 ^{\circ}$,  
thin solid one is $r=20 R_{*}$ and $\alpha'= 45 ^{\circ}$, and  
thin two-dashed one is $r=20 R_{*}$ and $\alpha'= 90 ^{\circ}$.  
The dependency of the polarization on $\tau_{0}$ become small 
for large $\tau_{0}$ ($\gtrsim 2$).
}
\label{fig10}
\end{figure}

In this section, we give a rough estimation for the size
and mass of dust puff responsible for the amplitude of the 
transient polarization ($p_{\rm var}\simeq 0.5$\%)
using the Monte-Carlo method.
In Table \ref{tbl-2} we summarized the results at 5500 \AA\ for both 
$r=2 R_{\rm *}$ and $20 R_{\rm *}$ cases.
Since the wavelength dependence is expected to be small
for the dust model, they can be considered as typical values 
in the optical region.
It is obvious that the models of $\alpha' = 135 ^{\circ}$ 
(backward scattering) cannot produce $p > 0.5$,
where $\alpha'$ is the scattering angle at the cloud center.
Figure \ref{fig10} shows the relations of $\tau _{0}$ vs. $\phi$
satisfying polarization of $0.5$\%.
To obtain the results of $p \simeq 0.5$ \%, we need (i) 
$\tau_{0}\gtrsim 0.5$--$1$ and $\phi \gtrsim 40\arcdeg$ for 
$r=2 R_{\rm *}$, or (ii) $\tau_{0}\gtrsim 0.5$ and 
$\phi \gtrsim 20\arcdeg$ for $r=20 R_{\rm *}$.
The cloud at larger distance can produce $p=0.5$ \% with smaller
optical thickness and smaller opening angle because the 
depolarization effect due to finite solid angle of the
photosphere become small.
However, larger mass is needed for large $r$.
We can estimate the total mass of the dust puff as
\begin{eqnarray}
M_{\rm d} & = & \frac{4\pi}{3}\rho \langle a^3\rangle
              \frac{\tau _{0}}{\langle \sigma _{\rm ext}\rangle } 
              r^2 \Omega \\
          &\simeq & 2.7 \times 10^{-12} 
           \left( \frac{\tau _{0}}{1.0} \right)
           \left( \frac{r}{R_{\rm *}} \right)^2
           \left( 1 - \cos \frac{\phi}{2} \right) \quad M_{\sun},
          \label{eqn2}
\end{eqnarray}
where $\rho$ is the bulk density of dust grains, 
$\langle a^3\rangle$ and $\langle \sigma _{\rm ext}\rangle$ are
the averaged cubic radius and averaged cross section of extinction
per a dust particle, respectively, and $\Omega$ is solid angle
of the puff seen from the star.
In Equation \ref{eqn2} we adopted 
$\langle a^3\rangle = 3.80 \times 10^{-18}$ cm$^{3}$ and
$\langle \sigma _{\rm ext}\rangle =1.42 \times 10^{-12}$ 
cm$^{2}$ \citep{whi79},
$\rho=2.2$ g cm$^{-3}$ \citep{eva93}.
We have 
$M_{\rm d} \simeq 9 \times 10^{-13}\tau_{0}\ M_{\sun}$ 
for the $r=2 R_{\rm *}$ case ($\phi = 48^{\circ}$), and 
$3 \times 10^{-11}\tau_{0}\ M_{\sun}$ for the $r=20 R_{*}$
case ($\phi = 28^{\circ}$).

\clearpage

\clearpage

\begin{deluxetable}{rrrcccrcccccr}
\tabletypesize{\scriptsize}
\rotate
\tablecolumns{13}
\tablecaption{Polarimetry of R CrB. \label{tbl-1}}
\tablewidth{0pt}
\tablehead{
\colhead{Date} & \colhead{JD} & \colhead{$\Delta m_{\rm V}$\tablenotemark{a}} &
\colhead{Tel.\tablenotemark{b}}   &
\colhead{Diaph.\tablenotemark{c}} &
\colhead{CCD\tablenotemark{d}} & \colhead{Exp.} & 
\colhead{$C_{2}\ 5165$\tablenotemark{e}} &
\multicolumn{2}{c}{Observed polarization\tablenotemark{f}} &
\multicolumn{2}{c}{Variable component\tablenotemark{f}} &
\colhead{Pol. flux\tablenotemark{g}} \\
\cline{9-10} \cline{11-12}
\colhead{{\it yyyymmdd}} & \colhead{2450000+} &
\colhead{$6.11+$} & \colhead{}  &
\colhead{} & \colhead{} & \colhead{sec} & 
\colhead{(EW) \AA} &
\colhead{$p_{\rm V}$ (\%)} & \colhead{$\theta_{\rm V}$ (deg.)} &
\colhead{$p_{\rm V}$ (\%)} & \colhead{$\theta_{\rm V}$ (deg.)} &
\colhead{$10^{-5}F_{\rm *}$}
}
\startdata
19980128 & 0842.32 & $-0.06$ & D36 & D1 & T &  1200 &   --- & $ 0.24\pm 0.01$ & $100.8\pm 1.8$ & $ 0.06\pm 0.02$ & $124.3\pm   6.5$& $ 64\pm 16$\\
19980325 & 0898.20 & $-0.00$ & D36 & D2 & T &  1440 &  $18$ & $ 0.13\pm 0.02$ & $ 96.7\pm 4.9$ & $ 0.07\pm 0.02$ & $181.4\pm   9.1$& $ 73\pm 22$\\
19980328 & 0901.23 & $-0.03$ & D36 & D1 & T &  1680 &   --- & $ 0.17\pm 0.02$ & $ 96.1\pm 3.3$ & $ 0.04\pm 0.02$ & $179.7\pm  13.4$& $ 41\pm 21$\\
19980329 & 0902.19 & $-0.08$ & D36 & D1 & T &   792 &   --- & $ 0.19\pm 0.03$ & $ 96.2\pm 4.1$ & $ 0.02\pm 0.03$ & $168.9\pm  46.3$& $ 19\pm 32$\\
19980522 & 0956.07 & $-0.12$ & D36 & D1 & T &  1260 &   --- & $ 0.18\pm 0.02$ & $ 98.3\pm 3.3$ & $ 0.04\pm 0.02$ & $168.0\pm  15.0$& $ 42\pm 21$\\
19980824 & 1050.01 & $+0.40$ & O74 & D1 & T &   960 &   --- & $ 0.22\pm 0.01$ & $ 96.9\pm 1.3$ & $ 0.02\pm 0.01$ & $117.0\pm  11.8$& $ 16\pm  7$\\
19990222 & 1232.23 & $+1.82$ & D36 & D1 & T &  3200 &   --- & $ 0.08\pm 0.02$ & $ 87.4\pm 7.4$ & $ 0.13\pm 0.02$ & $189.2\pm   4.3$& $ 25\pm  4$\\
19990223 & 1233.22 & $+1.80$ & D36 & D1 & T &  2880 &   --- & $ 0.10\pm 0.02$ & $ 88.8\pm 5.3$ & $ 0.11\pm 0.02$ & $190.1\pm   4.6$& $ 21\pm  3$\\
19990225 & 1235.23 & $+1.67$ & D36 & D1 & T &  2880 &   --- & $ 0.11\pm 0.01$ & $ 86.5\pm 3.5$ & $ 0.11\pm 0.01$ & $193.1\pm   3.5$& $ 23\pm  3$\\
19990227 & 1237.19 & $+1.60$ & D36 & D1 & T &  4200 &   --- & $ 0.14\pm 0.01$ & $ 90.8\pm 2.9$ & $ 0.07\pm 0.01$ & $193.0\pm   5.7$& $ 16\pm  3$\\
19990316 & 1254.18 & $+0.85$ & D36 & D1 & T &  1800 &   --- & $ 0.17\pm 0.01$ & $112.3\pm 1.4$ & $ 0.12\pm 0.01$ & $156.0\pm   2.1$& $ 56\pm  4$\\
19990416 & 1285.16 & $+0.38$ & D36 & D1 & T &  1800 &   --- & $ 0.19\pm 0.02$ & $101.2\pm 3.3$ & $ 0.05\pm 0.02$ & $150.5\pm  14.2$& $ 32\pm 15$\\
19990501 & 1300.06 & $+0.11$ & D36 & D1 & T &  1680 &   --- & $ 0.24\pm 0.01$ & $102.0\pm 1.7$ & $ 0.06\pm 0.01$ & $128.1\pm   5.8$& $ 59\pm 13$\\
19990510 & 1309.19 & $+0.09$ & O74 & D1 & T &  1440 &   --- & $ 0.25\pm 0.02$ & $101.2\pm 1.6$ & $ 0.07\pm 0.01$ & $122.8\pm   6.4$& $ 61\pm 13$\\
19990517 & 1316.21 & $+0.11$ & O74 & D1 & T &  1680 &   --- & $ 0.18\pm 0.02$ & $100.4\pm 3.2$ & $ 0.05\pm 0.02$ & $159.8\pm  12.1$& $ 41\pm 17$\\
19990808 & 1399.01 & $+1.09$ & D36 & D1 & S &  1500 &   --- & $ 0.32\pm 0.02$ & $108.9\pm 1.8$ & $ 0.17\pm 0.02$ & $126.7\pm   3.0$& $ 60\pm  8$\\
19991202 & 1515.34 & $+1.87$ & D36 & D1 & S &  1200 &   --- & $ 0.18\pm 0.04$ & $ 55.2\pm 5.9$ & $ 0.25\pm 0.04$ & $ 28.1\pm   4.4$& $ 43\pm  7$\\
19991203 & 1516.34 & $+1.87$ & D36 & D1 & S &  1200 &   --- & $ 0.29\pm 0.03$ & $ 48.2\pm 4.0$ & $ 0.37\pm 0.04$ & $ 31.1\pm   3.1$& $ 64\pm  6$\\
19991207 & 1520.33 & $+2.42$ & D36 & D1 & S &  1920 &   --- & $ 0.55\pm 0.02$ & $ 60.3\pm 1.2$ & $ 0.51\pm 0.02$ & $ 49.3\pm   1.3$& $ 56\pm  2$\\
20000101 & 1545.33 & $+1.16$ & D36 & D1 & S &  2400 &   --- & $ 0.26\pm 0.03$ & $ 96.6\pm 3.7$ & $ 0.05\pm 0.03$ & $103.4\pm  17.2$& $ 18\pm 10$\\
20000127 & 1571.30 & $+0.72$ & D36 & D1 & S &  1560 &   --- & $ 0.30\pm 0.03$ & $100.4\pm 2.3$ & $ 0.11\pm 0.03$ & $111.4\pm   6.7$& $ 55\pm 13$\\
20000203 & 1578.30 & $+0.64$ & D36 & D1 & S &   640 &   --- & $ 0.33\pm 0.04$ & $ 89.4\pm 3.5$ & $ 0.14\pm 0.04$ & $ 81.1\pm   8.6$& $ 78\pm 24$\\
20000204 & 1579.26 & $+0.55$ & D36 & D1 & S &  1680 &   --- & $ 0.22\pm 0.02$ & $ 97.1\pm 2.4$ & $ 0.03\pm 0.02$ & $117.4\pm  21.8$& $ 16\pm 12$\\
20000229 & 1604.19 & $+0.42$ & D36 & D1 & S &  2176 &   --- & $ 0.13\pm 0.02$ & $103.1\pm 5.2$ & $ 0.09\pm 0.02$ & $171.8\pm   8.2$& $ 60\pm 14$\\
20000306 & 1610.22 & $+0.53$ & D36 & D1 & S &  2560 &   --- & $ 0.18\pm 0.03$ & $105.5\pm 5.7$ & $ 0.08\pm 0.04$ & $154.1\pm  13.1$& $ 48\pm 22$\\
20000329 & 1633.15 & $+0.54$ & D36 & D1 & S &  2080 &   --- & $ 0.14\pm 0.02$ & $ 78.4\pm 4.2$ & $ 0.12\pm 0.02$ & $ 25.6\pm   5.0$& $ 73\pm 13$\\
20001207 & 1886.35 & $+1.64$ & O36 & D1 & S &  1500 &   --- & $ 0.64\pm 0.04$ & $ 88.4\pm 2.0$ & $ 0.44\pm 0.04$ & $ 85.5\pm   2.8$& $102\pm 10$\\
20001214 & 1893.35 & $+4.18$ & O36 & D1 & S &  2800 &   --- & $ 6.81\pm 0.31$ & $102.0\pm 1.4$ & $ 6.61\pm 0.31$ & $102.2\pm   1.4$& $138\pm  6$\\
20001215 & 1894.35 & $+4.52$ & O36 & D1 & S &  2400 &   --- & $ 9.98\pm 0.42$ & $ 98.4\pm 1.2$ & $ 9.78\pm 0.42$ & $ 98.5\pm   1.3$& $155\pm  7$\\
20010130 & 1940.30 & $+1.13$ & O74 & D2 & T &  5600 &  $20$ & $ 0.29\pm 0.03$ & $107.0\pm 2.9$ & $ 0.13\pm 0.03$ & $127.1\pm   6.2$& $ 47\pm 11$\\
20010202 & 1943.32 & $+0.84$ & O74 & D1 & T &  2240 &   --- & $ 0.15\pm 0.03$ & $132.9\pm 5.6$ & $ 0.23\pm 0.03$ & $164.2\pm   3.8$& $108\pm 14$\\
20010216 & 1957.34 & $+0.26$ & O36 & D2 & S &  1680 &  $30$ & $ 0.13\pm 0.01$ & $104.9\pm 3.0$ & $ 0.10\pm 0.01$ & $171.5\pm   3.9$& $ 74\pm 11$\\
20010226 & 1967.29 & $+0.15$ & O36 & D2 & S &  2400 &  $23$ & $ 0.22\pm 0.02$ & $ 85.6\pm 2.2$ & $ 0.07\pm 0.02$ & $ 51.7\pm   6.9$& $ 64\pm 16$\\
20010309 & 1978.30 & $+0.13$ & O36 & D2 & S &  1800 &  $23$ & $ 0.74\pm 0.03$ & $ 92.6\pm 1.1$ & $ 0.53\pm 0.03$ & $ 91.7\pm   1.6$& $487\pm 28$\\
20010311 & 1980.17 & $+0.07$ & O36 & D1 & S &  3840 &   --- & $ 0.71\pm 0.03$ & $ 85.8\pm 1.2$ & $ 0.52\pm 0.03$ & $ 82.2\pm   1.6$& $475\pm 25$\\
20010312 & 1981.24 & $+0.11$ & O36 & D2 & S &  3640 &  $25$ & $ 0.50\pm 0.03$ & $ 92.1\pm 1.6$ & $ 0.30\pm 0.03$ & $ 90.2\pm   2.6$& $269\pm 26$\\
20010313 & 1982.22 & $+0.08$ & O36 & D2 & S &  2000 &  $25$ & $ 0.15\pm 0.01$ & $ 98.3\pm 2.3$ & $ 0.06\pm 0.01$ & $176.6\pm   5.5$& $ 57\pm 11$\\
20010418 & 2018.09 & $+0.03$ & O36 & D2 & S &  1440 &  $24$ & $ 0.13\pm 0.02$ & $ 99.6\pm 5.0$ & $ 0.08\pm 0.02$ & $177.0\pm   8.3$& $ 79\pm 23$\\
20010809 & 2131.05 & $+0.00$ & O36 & D2 & S &   900 &  $22$ & $ 0.66\pm 0.06$ & $ 74.8\pm 2.5$ & $ 0.52\pm 0.06$ & $ 67.4\pm   3.1$& $520\pm 56$\\
20010811 & 2133.06 & $+0.02$ & O36 & D2 & S &  1200 &  $24$ & $ 0.30\pm 0.02$ & $ 91.7\pm 2.0$ & $ 0.10\pm 0.02$ & $ 85.3\pm   5.9$& $102\pm 22$\\
20010904 & 2156.97 & $-0.04$ & O36 & D2 & S &  1080 &  $28$ & $ 0.24\pm 0.02$ & $ 94.5\pm 2.5$ & $ 0.04\pm 0.02$ & $ 92.5\pm  15.6$& $ 39\pm 23$\\
20010908 & 2161.04 & $+0.05$ & O36 & D2 & S &   960 &  $29$ & $ 0.21\pm 0.03$ & $ 88.4\pm 4.6$ & $ 0.05\pm 0.03$ & $ 51.3\pm  20.4$& $ 44\pm 31$\\
20010909 & 2162.03 & $+0.06$ & O36 & D2 & S &  1080 &  $29$ & $ 0.12\pm 0.03$ & $ 98.4\pm 6.4$ & $ 0.09\pm 0.03$ & $179.9\pm   8.9$& $ 79\pm 26$\\
20010926 & 2178.95 & $-0.02$ & O36 & D2 & S &  1200 &  $27$ & $ 0.18\pm 0.02$ & $ 94.8\pm 2.9$ & $ 0.03\pm 0.02$ & $185.4\pm  17.1$& $ 30\pm 19$\\
20011030 & 2212.90 & $-0.01$ & O36 & D2 & S &   960 &  $29$ & $ 0.16\pm 0.01$ & $ 97.4\pm 2.7$ & $ 0.05\pm 0.01$ & $176.1\pm   9.1$& $ 47\pm 14$\\
20011031 & 2213.89 & $+0.00$ & O36 & D2 & S &  1120 &  $29$ & $ 0.27\pm 0.02$ & $102.9\pm 2.3$ & $ 0.09\pm 0.02$ & $122.2\pm   6.7$& $ 91\pm 22$\\
20011201 & 2245.36 & $+0.12$ & O36 & D2 & S &  1600 &  $22$ & $ 0.31\pm 0.01$ & $ 83.9\pm 1.2$ & $ 0.14\pm 0.01$ & $ 67.8\pm   2.6$& $131\pm 12$\\
20011202 & 2246.36 & $+0.07$ & O36 & D2 & S &  1200 &  $20$ & $ 0.29\pm 0.02$ & $ 93.9\pm 2.6$ & $ 0.08\pm 0.02$ & $ 91.5\pm   9.0$& $ 75\pm 19$\\
20011227 & 2271.35 & $-0.05$ & O36 & D2 & S &  3000 &  $25$ & $ 0.28\pm 0.01$ & $ 89.8\pm 1.3$ & $ 0.08\pm 0.01$ & $ 76.6\pm   4.2$& $ 81\pm 12$\\
20020122 & 2297.36 & $-0.04$ & O36 & D2 & S &  1200 &  $18$ & $ 0.18\pm 0.01$ & $106.9\pm 1.6$ & $ 0.09\pm 0.01$ & $154.7\pm   3.2$& $ 85\pm 10$\\
20020124 & 2299.37 & $-0.05$ & O36 & D2 & S &  1080 &  $19$ & $ 0.14\pm 0.01$ & $111.9\pm 2.4$ & $ 0.12\pm 0.01$ & $164.6\pm   2.8$& $120\pm 11$\\
20020131 & 2306.31 & $-0.03$ & O36 & D2 & S &  1800 &  $25$ & $ 0.19\pm 0.01$ & $ 94.4\pm 2.0$ & $ 0.02\pm 0.01$ & $189.9\pm  22.6$& $ 17\pm 14$\\
20020206 & 2312.28 & $+0.01$ & O36 & D2 & S &  2160 &  $29$ & $ 0.14\pm 0.01$ & $ 99.3\pm 2.4$ & $ 0.07\pm 0.01$ & $175.2\pm   5.1$& $ 68\pm 11$\\
20020207 & 2313.26 & $-0.01$ & O36 & D2 & S &  2160 &  $28$ & $ 0.17\pm 0.01$ & $ 97.6\pm 1.6$ & $ 0.04\pm 0.01$ & $174.6\pm   5.8$& $ 44\pm 10$\\
20020224 & 2330.27 & $-0.12$ & O74 & D2 & S &   360 &  $26$ & $ 0.22\pm 0.02$ & $105.1\pm 2.6$ & $ 0.08\pm 0.02$ & $140.1\pm   7.8$& $ 84\pm 21$\\
20020401 & 2366.31 & $-0.01$ & O36 & D2 & S &  1260 &  $22$ & $ 0.22\pm 0.02$ & $ 98.6\pm 1.9$ & $ 0.03\pm 0.02$ & $125.0\pm  12.9$& $ 33\pm 15$\\
20020403 & 2368.22 & $-0.01$ & O36 & D2 & S &  2700 &  $23$ & $ 0.27\pm 0.01$ & $ 93.8\pm 0.6$ & $ 0.07\pm 0.01$ & $ 90.6\pm   2.6$& $ 67\pm  7$\\
20020501 & 2396.11 & $+0.04$ & O36 & D2 & S &  2880 &  $29$ & $ 0.22\pm 0.01$ & $ 82.4\pm 1.1$ & $ 0.09\pm 0.01$ & $ 47.6\pm   2.5$& $ 92\pm  8$\\
20020803 & 2489.97 & $-0.02$ & O36 & D2 & S &   900 &  $14$ & $ 0.21\pm 0.02$ & $ 93.9\pm 2.6$ & $ 0.01\pm 0.02$ & $ 24.5\pm  51.1$& $ 10\pm 18$\\
20020804 & 2490.98 & $-0.01$ & O36 & D2 & S &   900 &  $25$ & $ 0.23\pm 0.01$ & $ 90.7\pm 1.0$ & $ 0.04\pm 0.01$ & $ 67.4\pm   6.1$& $ 40\pm  9$\\
20020805 & 2492.04 & $-0.01$ & O36 & D2 & S &   900 &  $24$ & $ 0.22\pm 0.02$ & $ 91.4\pm 2.1$ & $ 0.03\pm 0.02$ & $ 62.4\pm  16.6$& $ 29\pm 16$\\
20020806 & 2492.99 & $-0.00$ & O36 & D2 & S &  1260 &  $25$ & $ 0.23\pm 0.01$ & $ 87.3\pm 1.5$ & $ 0.06\pm 0.01$ & $ 58.7\pm   6.1$& $ 64\pm 13$\\
20020807 & 2493.96 & $+0.01$ & O36 & D2 & S &   972 &  $25$ & $ 0.23\pm 0.01$ & $ 93.1\pm 1.0$ & $ 0.03\pm 0.01$ & $ 80.6\pm   7.7$& $ 30\pm  8$\\
20030204 & 2675.33 & $+0.07$ & O36 & D2 & S &  1800 &  $30$ & $ 0.23\pm 0.01$ & $ 84.4\pm 1.5$ & $ 0.08\pm 0.01$ & $ 53.8\pm   3.8$& $ 76\pm 11$\\
20030205 & 2676.31 & $+0.06$ & O36 & D2 & S &  1440 &  $29$ & $ 0.22\pm 0.01$ & $ 79.7\pm 1.4$ & $ 0.11\pm 0.01$ & $ 46.4\pm   2.5$& $103\pm 10$\\
20030206 & 2677.25 & $+0.04$ & O36 & D2 & S &  1600 &  $29$ & $ 0.21\pm 0.01$ & $ 84.8\pm 1.0$ & $ 0.07\pm 0.01$ & $ 45.2\pm   3.2$& $ 72\pm  7$\\
20030211 & 2682.33 & $+0.66$ & O36 & D2 & S &  3200 &  $27$ & $ 0.21\pm 0.02$ & $ 97.2\pm 1.8$ & $ 0.02\pm 0.01$ & $132.8\pm  24.6$& $  9\pm  7$\\
20030212 & 2683.29 & $+0.70$ & O36 & D2 & S &  2880 &  $27$ & $ 0.20\pm 0.01$ & $101.1\pm 1.6$ & $ 0.05\pm 0.01$ & $148.0\pm   6.0$& $ 24\pm  6$\\
20030213 & 2684.32 & $+0.76$ & O36 & D2 & S &  2400 &  $25$ & $ 0.30\pm 0.01$ & $103.7\pm 0.9$ & $ 0.12\pm 0.01$ & $118.8\pm   2.2$& $ 59\pm  5$\\
20030214 & 2685.31 & $+0.95$ & O36 & D2 & S &  4000 &  $24$ & $ 0.30\pm 0.01$ & $104.4\pm 0.8$ & $ 0.12\pm 0.01$ & $121.0\pm   1.9$& $ 54\pm  3$\\
20030216 & 2687.34 & $+1.48$ & O36 & D2 & S &  2400 &  $23$ & $ 0.34\pm 0.01$ & $120.4\pm 0.8$ & $ 0.27\pm 0.01$ & $138.9\pm   1.1$& $ 67\pm  3$\\
20030217 & 2688.30 & $+1.69$ & O36 & D2 & S &  2400 &  $22$ & $ 0.42\pm 0.04$ & $126.5\pm 2.6$ & $ 0.38\pm 0.04$ & $141.0\pm   2.8$& $ 79\pm  9$\\
20030218 & 2689.34 & $+2.08$ & O36 & D2 & S &  2400 &  $24$ & $ 0.67\pm 0.02$ & $127.6\pm 0.8$ & $ 0.62\pm 0.02$ & $136.4\pm   0.9$& $ 89\pm  3$\\
20030220 & 2691.28 & $+3.09$ & O36 & D2 & S &  3200 &  $32$ & $ 0.97\pm 0.02$ & $141.2\pm 0.6$ & $ 1.00\pm 0.02$ & $147.1\pm   0.5$& $ 57\pm  1$\\
20030319 & 2718.26 & $+5.44$ & O36 & D2 & S &  9600 &  $10$ & $ 1.64\pm 0.06$ & $ 32.5\pm 0.8$ & $ 1.77\pm 0.06$ & $ 29.7\pm   0.8$& $ 12\pm 0.4$\\
20030320 & 2719.23 & $+5.17$ & O36 & D2 & S & 10800 &  $13$ & $ 1.55\pm 0.04$ & $ 35.5\pm 0.7$ & $ 1.66\pm 0.04$ & $ 32.3\pm   0.7$& $ 14\pm 0.3$\\
20030322 & 2721.19 & $+4.83$ & O36 & D2 & S &  9600 &  $10$ & $ 1.23\pm 0.04$ & $ 40.6\pm 0.8$ & $ 1.31\pm 0.04$ & $ 36.3\pm   0.7$& $ 16\pm 0.5$\\
20030522 & 2782.12 & $+0.02$ & O36 & D2 & T &  4140 &  $24$ & $ 0.26\pm 0.02$ & $105.7\pm 2.6$ & $ 0.10\pm 0.02$ & $129.3\pm   5.9$& $103\pm 24$\\
20030523 & 2783.02 & $+0.03$ & O36 & D2 & T &  3200 &  $24$ & $ 0.27\pm 0.01$ & $110.8\pm 1.5$ & $ 0.14\pm 0.02$ & $135.9\pm   2.6$& $141\pm 15$\\
20030725 & 2845.97 & $+0.04$ & O36 & D2 & S &   700 &  $44$ & $ 0.24\pm 0.01$ & $ 95.7\pm 1.6$ & $ 0.04\pm 0.01$ & $100.4\pm  10.4$& $ 38\pm 12$\\
20030727 & 2848.09 & $+0.05$ & O36 & D2 & S &   552 &  $41$ & $ 0.21\pm 0.01$ & $ 94.9\pm 2.1$ & $ 0.00\pm 0.01$ & $164.1\pm 550.3$& $  1\pm 124$\\
20030926 & 2908.96 & $-0.07$ & O36 & D2 & S &  1800 &  $18$ & $ 0.24\pm 0.01$ & $ 90.4\pm 1.5$ & $ 0.05\pm 0.01$ & $ 67.8\pm   7.6$& $ 50\pm 13$\\
20030928 & 2910.92 & $-0.05$ & O36 & D2 & S &  1080 &  $20$ & $ 0.24\pm 0.02$ & $ 92.5\pm 1.8$ & $ 0.04\pm 0.02$ & $ 79.5\pm  11.3$& $ 38\pm 15$\\
20030929 & 2911.92 & $-0.04$ & O36 & D2 & S &   960 &  $18$ & $ 0.24\pm 0.01$ & $ 85.6\pm 1.6$ & $ 0.08\pm 0.01$ & $ 57.0\pm   3.9$& $ 77\pm 13$\\
\enddata

\tablenotetext{a}{Visual magnitude averaged for
all data within $\pm 2.5$ days in VSNET database.}
\tablenotetext{b}{D36, O36 and O74 denote 0.91 m at Dodaira, 
0.91 m at Okayama and 1.88 m at Okayama, respectively.}
\tablenotetext{c}{D1: 1.4 mm$\phi$ circle, D2: 0.2 mm $\times$ 1.4 mm rectangle.}
\tablenotetext{d}{T and S denote TI CCD and SITe CCD, respectively.}
\tablenotetext{e}{The equivalent width of the C$_{2}\ (0-0)$ 5165 \AA\
 band is measured between 4970 \AA\ and 5200 \AA\ for D2 observations.}
\tablenotetext{f}{Observed polarization is considered as a vectorial 
summation of a constant component and a variable component (see text).}
\tablenotetext{g}{Polarized flux $p_{\rm var} \times 10^{-0.4 \Delta m_{\rm V}}\ F_{\rm *}$}

\end{deluxetable}

\clearpage
\begin{table*}
{\scriptsize
\begin{center}
\begin{tabular}{lllcccc}

\hline
\hline
$r$ & $\alpha$ & $\phi$ & 
\multicolumn{4}{c}{\% polarization} \\
\cline{4-7}
& & & 
$\tau_{0}=0.3$ & $\tau_{0}=1.0$ & $\tau_{0}=3.0$ & $\tau_{0}=5.0$ \\
\hline
$2\ R_{\rm *}$ &
$45\arcdeg$ & $20\arcdeg$ & 0.035 & 0.081 & 0.118 & 0.122\\
      &     & $30\arcdeg$ & 0.074 & 0.175 & 0.250 & 0.259\\
      &     & $45\arcdeg$ & 0.151 & 0.351 & 0.497 & 0.511\\
      &     & $60\arcdeg$ & 0.232 & 0.533 & 0.742 & 0.768\\
\hline  
      &      
$90\arcdeg$ & $20\arcdeg$ & 0.042 & 0.100 & 0.144 & 0.148\\
      &     & $30\arcdeg$ & 0.092 & 0.218 & 0.311 & 0.321\\
      &     & $45\arcdeg$ & 0.191 & 0.453 & 0.640 & 0.663\\
      &     & $60\arcdeg$ & 0.306 & 0.718 & 1.014 & 1.051\\
\hline 
      &       
$135\arcdeg$ & $20\arcdeg$ & 0.011 & 0.026 & 0.038 & 0.039\\
      &      & $30\arcdeg$ & 0.024 & 0.058 & 0.083 & 0.087\\
      &      & $45\arcdeg$ & 0.052 & 0.122 & 0.174 & 0.180\\
      &      & $60\arcdeg$ & 0.086 & 0.200 & 0.287 & 0.297\\
\hline
\hline         
$20\ R_{\rm *}$ &
$45\arcdeg$ & $10\arcdeg$ & 0.030 & 0.073 & 0.110 & 0.115\\
      &     & $20\arcdeg$ & 0.116 & 0.278 & 0.415 & 0.434\\
      &     & $30\arcdeg$ & 0.243 & 0.579 & 0.856 & 0.892\\
      &     & $45\arcdeg$ & 0.471 & 1.089 & 1.588 & 1.634\\
\hline 
      &       
$90\arcdeg$ & $10\arcdeg$ & 0.028 & 0.069 & 0.104 & 0.109\\
      &     & $20\arcdeg$ & 0.111 & 0.269 & 0.404 & 0.422\\
      &     & $30\arcdeg$ & 0.239 & 0.577 & 0.865 & 0.901\\
      &     & $45\arcdeg$ & 0.489 & 1.169 & 1.732 & 1.804\\
\hline 
      &       
$135\arcdeg$ & $10\arcdeg$ & 0.005 & 0.012 & 0.018 & 0.019\\
      &      & $20\arcdeg$ & 0.020 & 0.049 & 0.073 & 0.077\\
      &      & $30\arcdeg$ & 0.045 & 0.109 & 0.164 & 0.171\\
      &      & $45\arcdeg$ & 0.100 & 0.242 & 0.361 & 0.376\\
\hline         
\end{tabular}
\caption{Monte Carlo simulation of polarization
due to scattering by dust puff\label{tbl-2}}
\end{center}
}
\end{table*}


\end{document}